\documentclass[twocolumn]{aastex62}

\graphicspath{{./}{figures/}}

\submitjournal{ApJL}

\shorttitle{Joint constraints on Galactic diffuse neutrino emission from ANTARES and IceCube}

\usepackage{amsmath}
\usepackage{graphicx}
\usepackage{multirow}
\usepackage{units}
\usepackage{booktabs}

\newcommand{\Fermi}{\emph{Fermi}}

\begin{document}

\title{Joint constraints on Galactic diffuse neutrino emission from the ANTARES and IceCube neutrino telescopes}

\collaboration{ANTARES collaboration}
\author{A.~Albert}
\affiliation{\scriptsize{Universit\'e de Strasbourg, CNRS,  IPHC UMR 7178, F-67000 Strasbourg, France}\email{antares.spokesperson@in2p3.fr}}
\author{M.~Andr\'e}
\affiliation{\scriptsize{Technical University of Catalonia, Laboratory of Applied Bioacoustics, Rambla Exposici\'o, 08800 Vilanova i la Geltr\'u, Barcelona, Spain}}
\author{M.~Anghinolfi}
\affiliation{\scriptsize{INFN - Sezione di Genova, Via Dodecaneso 33, 16146 Genova, Italy}}
\author{M.~Ardid}
\affiliation{\scriptsize{Institut d'Investigaci\'o per a la Gesti\'o Integrada de les Zones Costaneres (IGIC) - Universitat Polit\`ecnica de Val\`encia. C/  Paranimf 1, 46730 Gandia, Spain}}
\author{J.-J.~Aubert}
\affiliation{\scriptsize{Aix Marseille Univ, CNRS/IN2P3, CPPM, Marseille, France}}
\author{J.~Aublin}
\affiliation{\scriptsize{APC, Univ Paris Diderot, CNRS/IN2P3, CEA/Irfu, Obs de Paris, Sorbonne Paris Cit\'e, France}}
\author{T.~Avgitas}
\affiliation{\scriptsize{APC, Univ Paris Diderot, CNRS/IN2P3, CEA/Irfu, Obs de Paris, Sorbonne Paris Cit\'e, France}}
\author{B.~Baret}
\affiliation{\scriptsize{APC, Univ Paris Diderot, CNRS/IN2P3, CEA/Irfu, Obs de Paris, Sorbonne Paris Cit\'e, France}}
\author{J.~Barrios-Mart\'{\i}}
\affiliation{\scriptsize{IFIC - Instituto de F\'\i sica Corpuscular (CSIC - Universitat de Val\`encia) c/ Catedr\'atico Jos\'e Beltr\'an, 2 E-46980 Paterna, Valencia, Spain}}
\author{S.~Basa}
\affiliation{\scriptsize{LAM - Laboratoire d'Astrophysique de Marseille, P\^ole de l'\'Etoile Site de Ch\^ateau-Gombert, rue Fr\'ed\'eric Joliot-Curie 38,  13388 Marseille Cedex 13, France}}
% ...


\collaboration{IceCube Collaboration}
\author{M.~G.~Aartsen}
\affiliation{Dept.~of Physics and Astronomy, University of Canterbury, Private Bag 4800, Christchurch, New Zealand\email{analysis@icecube.wisc.edu}}
\author{M.~Ackermann}
\affiliation{DESY, D-15738 Zeuthen, Germany}
\author{J.~Adams}
\affiliation{Dept.~of Physics and Astronomy, University of Canterbury, Private Bag 4800, Christchurch, New Zealand}
\author{J.~A.~Aguilar}
\affiliation{Universit\'e Libre de Bruxelles, Science Faculty CP230, B-1050 Brussels, Belgium}
\author{M.~Ahlers}
\affiliation{Niels Bohr Institute, University of Copenhagen, DK-2100 Copenhagen, Denmark}
\author{M.~Ahrens}
\affiliation{Oskar Klein Centre and Dept.~of Physics, Stockholm University, SE-10691 Stockholm, Sweden}
\author{I.~Al~Samarai}
\affiliation{D\'epartement de physique nucl\'eaire et corpusculaire, Universit\'e de Gen\`eve, CH-1211 Gen\`eve, Switzerland}
\author{D.~Altmann}
\affiliation{Erlangen Centre for Astroparticle Physics, Friedrich-Alexander-Universit\"at Erlangen-N\"urnberg, D-91058 Erlangen, Germany}
\author{K.~Andeen}
\affiliation{Department of Physics, Marquette University, Milwaukee, WI, 53201, USA}
\author{T.~Anderson}
\affiliation{Dept.~of Physics, Pennsylvania State University, University Park, PA 16802, USA}
% ...

\collaboration{}
\author{D.~Gaggero}
\affiliation{\scriptsize{GRAPPA, University of Amsterdam, Science Park 904, 1098 XH Amsterdam, Netherlands}}
\author{D.~Grasso}
\affiliation{\scriptsize{INFN - Sezione di Pisa, Largo B. Pontecorvo 3, 56127 Pisa, Italy}}
\affiliation{\scriptsize{Dipartimento di Fisica dell'Universit\`a, Largo B. Pontecorvo 3, 56127 Pisa, Italy}}

\begin{abstract}
 
The existence of diffuse Galactic neutrino production is expected from cosmic ray interactions with Galactic gas and radiation fields. Thus, neutrinos are a unique messenger offering the opportunity to test the products of Galactic cosmic ray interactions up to energies of hundreds of \unit{TeV}. Here we present a search for this production using ten years of ANTARES track and shower data, as well as seven years of IceCube track data. The data are combined into a joint likelihood test for neutrino emission according to the KRA$_\gamma$ model assuming a \unit[5]{PeV} per nucleon Galactic cosmic ray cutoff. No significant excess is found. As a consequence, the limits presented in this work start constraining the model parameter space for Galactic cosmic ray production and transport.



\end{abstract}


\section{Introduction}

A diffuse Galactic neutrino emission is expected from cosmic ray (CR) interactions with interstellar gas and radiation fields. These interactions are also the dominant production mechanism of the diffuse high-energy $\gamma$-rays in the Galactic plane, which have been measured by the \Fermi-Large Area Telescope (\Fermi-LAT)~\citep{FermiTemplate}.

In the GALPROP-based \citep{GALPROPWebRun} conventional model of Galactic diffuse $\gamma$-ray production CRs are accelerated in a distribution of sources such as supernova remnants. They propagate diffusively in the interstellar medium producing $\gamma$-rays and neutrinos via interactions with the interstellar radiation field and interstellar gas.  The interstellar radiation field is weakly constrained by \Fermi-LAT $\gamma$-ray data and interstellar gas is constrained by both \Fermi-LAT $\gamma$-ray data and radio measurements of CO and HI line intensities.  The CR population model itself is normalised to local measurements taken at Earth. The GALPROP model parameters are tuned to achieve optimal agreement between \Fermi-LAT~\citep{FermiTemplate} data and the direction-dependent prediction given by integrating expected $\gamma$-ray yields along the line of sight from Earth. The neutral pion decay component estimated by the conventional model should be accompanied by a neutrino flux from charged pion decay.

\begin{figure}
 \centering
 \includegraphics[width=0.45\textwidth]{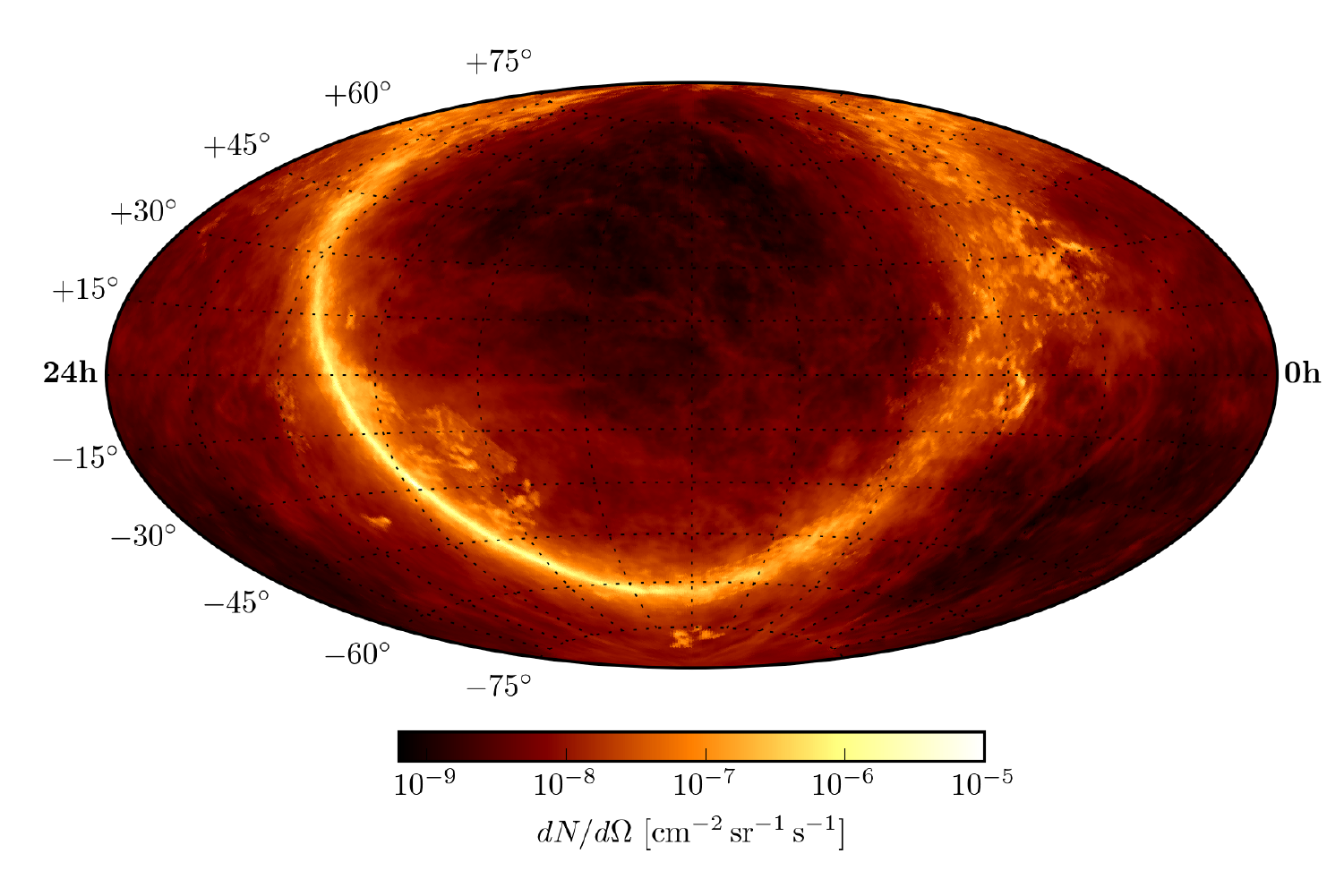}
 \caption{\label{fig:ModelFlux}Neutrino flux per unit of solid angle of the KRA$_\gamma^{5}$ model~\citep{KRAgamma}, shown as a function of direction in equatorial coordinates (Hammer projection).}
\end{figure}
The conventional model, however, under-predicts the $\gamma$-ray flux above \unit[10]{GeV} in the inner Galaxy~\citep{FermiTemplate}. The KRA$_\gamma$ models~\citep{KRAgamma, KRAgamma2, KRAgamma3} address this issue using a radially-dependent model for the CR diffusion coefficient and the advective wind. The primary CR spectrum assumed within the KRA$_\gamma$ models has an exponential cutoff at a certain energy.  
In order to bracket measurements by KASCADE~\citep{KASCADE} and KASCADE-Grande~\citep{KASCADEgrande}, respectively in the [\unit[100]{TeV}, \unit[100]{PeV}] and [\unit[10]{PeV}, \unit[2000]{PeV}] energy ranges, while maintaining agreement with proton and helium measurements by CREAM~\citep{CREAM_pHe}, cutoffs at 5 and \unit[50]{PeV} per nucleon are considered.  
The resulting models are referred to as KRA$_\gamma^{5}$ and KRA$_\gamma^{50}$, respectively. The direction dependence of the energy-integrated KRA$_\gamma^{5}$ neutrino flux prediction is shown in Figure~\ref{fig:ModelFlux}. Compared to the conventional model of the Galactic diffuse emission, the KRA$_\gamma$ models predict modified spectra and enhanced overall $\gamma$-ray and neutrino fluxes in the southern sky, especially in the central ridge where a hardening of the CR spectra is reproduced. Hence, neutrinos offer a unique opportunity to independently test the model assumptions of Galactic CR production and transport, accessing energies far beyond the reach of current $\gamma$-ray experiments.

The KRA$_\gamma$ predictions have already been tested separately with ANTARES~\citep{GalPlane_ANT} and IceCube~\citep{GalPlane_IC} data.  ANTARES and IceCube achieved sensitivities of $1.05\times\Phi_{\text{KRA}_\gamma^{50}}$ and $0.79\times\Phi_{\text{KRA}_\gamma^{50}}$, respectively; both analyses obtained 90\% confidence level (CL) upper limits of $1.2\times\Phi_{\text{KRA}_\gamma^{50}}$.  ANTARES additionally examined the \unit[5]{PeV} cutoff model, obtaining a sensitivity of $1.4\times\Phi_{\text{KRA}_\gamma^{5}}$ and an upper limit of $1.1\times\Phi_{\text{KRA}_\gamma^{5}}$ due to an under-fluctuation of the fitted signal flux in the track channel.

This paper presents a combination of these two maximum-likelihood analyses exploiting the advantageous field of view of ANTARES as well as the high statistics of IceCube.

\section{Detectors and Data samples}
The IceCube Neutrino Observatory~\citep{IC_Telescope} is located at the South Pole between 1.45 and \unit[2.45]{km} below the surface of the ice. It consists of 5160 photomultiplier tubes (PMTs) instrumenting one cubic kilometer of ice. The ANTARES neutrino telescope~\citep{ANT_Telescope} consists of 885 PMTs deployed in the Mediterranean sea, \unit[40]{km} off the coast of Toulon, France. It is installed at depths between \unit[2.01]{km} and \unit[2.47]{km} below sea level, instrumenting a volume of $\sim$\unit[0.01]{km$^3$}.

Neutrinos interacting with matter produce charged particles that generate Cherenkov light in the detectors. From the collected Cherenkov light, the energy and direction of the incoming neutrinos are reconstructed. A muon neutrino\footnote{In the following, particles also refer to the corresponding anti-particles.} undergoing a charged current interaction produces a muon that can travel large distances through the medium, leading to a \emph{track} event topology in the detector. Most other interactions produce a nearly spherical {\it shower} event topology. In this analysis, ANTARES events of both topologies are used, while only track events are taken from IceCube data.

The ANTARES event sample used in this work includes the one used in ~\citep{GalPlane_ANT} extended by the data collected in 2016. These data use the most recent offline-reconstructed dataset, incorporating dedicated calibrations of positioning, timing and  efficiency~\citep{position_calib}. The sample is taken from a total of 2780 days of detector livetime, over a total of ten calendar years. Part of the sample was collected with partially-completed detector configurations. Here, 218 shower(-like) events are selected, while 2.6 signal events are expected from the KRA$_\gamma^{5}$ model. For these signal events we have a median angular resolution of $2.4^\circ$.  The track selection includes 7,850 events, with 10.2 signal events expected having an angular resolution of $0.5^\circ$. The energy ranges including 90\% of signal events are [\unit[2.1]{TeV}, \unit[150]{TeV}] for showers and [\unit[360]{GeV}, \unit[130]{TeV}] for tracks.

The IceCube seven-year track selection used in this analysis is detailed by~\citet{IC_PS_7yrs}. It results in a total of 730,130 events with 191 events expected from the KRA$_\gamma^{5}$ model. The dataset was collected over a total of 2431 days of detector livetime, some of which took place during the construction phase of the detector. The IceCube signal events are expected to have median angular resolution of $0.8^\circ$. The energy range containing 90\% of the expected signal events is [\unit[390]{GeV}, \unit[110]{TeV}].

The energy range in which the combined analysis is valid is [\unit[90]{GeV}, \unit[300]{TeV}]. This range is defined as containing 90\% of the sensitivity. It is calculated by finding the low and high energy thresholds where removing simulated signal events outside these values worsens the sensitivity by 5\% each.

\section{Search Method}

The present analysis uses an unbinned likelihood ratio test. The likelihood functions for each sample | ANTARES tracks, ANTARES showers, and IceCube tracks | are defined as
\begin{equation}
 \begin{split}
  \mathcal{L}_\textrm{sig+bkg}(n_\textrm{sig}) = \prod\limits_{i}& \left[\frac{n_\textrm{sig}}{N}\cdot \mathcal{S}(E_i, \alpha_i, \delta_i)\right. \\
  &\left. + (1 - \frac{n_\textrm{sig}}{N})\cdot \mathcal{B}(E_i, \delta_i) \right],
 \end{split}
\end{equation}
where $N$ is the total number of events; $n_\textrm{sig}$ is the number of signal events; $\mathcal{S}$ is the signal probability density function (PDF) for an event $i$ at the equatorial coordinates $(\alpha_i, \delta_i)$ with energy $E_i$.  It is obtained from Monte Carlo simulations of the detectors with the model flux as input - it is proportional to the expected signal rate at a given reconstructed energy and direction.  $\mathcal{B}$ is the PDF of the background.


Minor differences in the original, separate ANTARES and IceCube PDF constructions are preserved in this work. For IceCube tracks, the background term $\mathcal{B}$ comes from the data with a correction for the signal contamination expected for $n_\textrm{sig}$ signal events~\citep{GalPlane_IC}. For the ANTARES samples, this is approximated by ignoring the signal correction term~\citep{GalPlane_ANT}. In addition, the IceCube signal PDF accounts for the estimated point spread function of each event, while average point spread functions are used for track and shower ANTARES events.

In order to account for the different acceptances of each sample as well as any bias in the fitted signal normalization, we forward-fold the signal flux $\Phi_\textrm{sig}$ into the individual likelihoods using a response function obtained from simulated pseudo-experiments.



\begin{figure}
 \centering
 \includegraphics[width=0.45\textwidth]{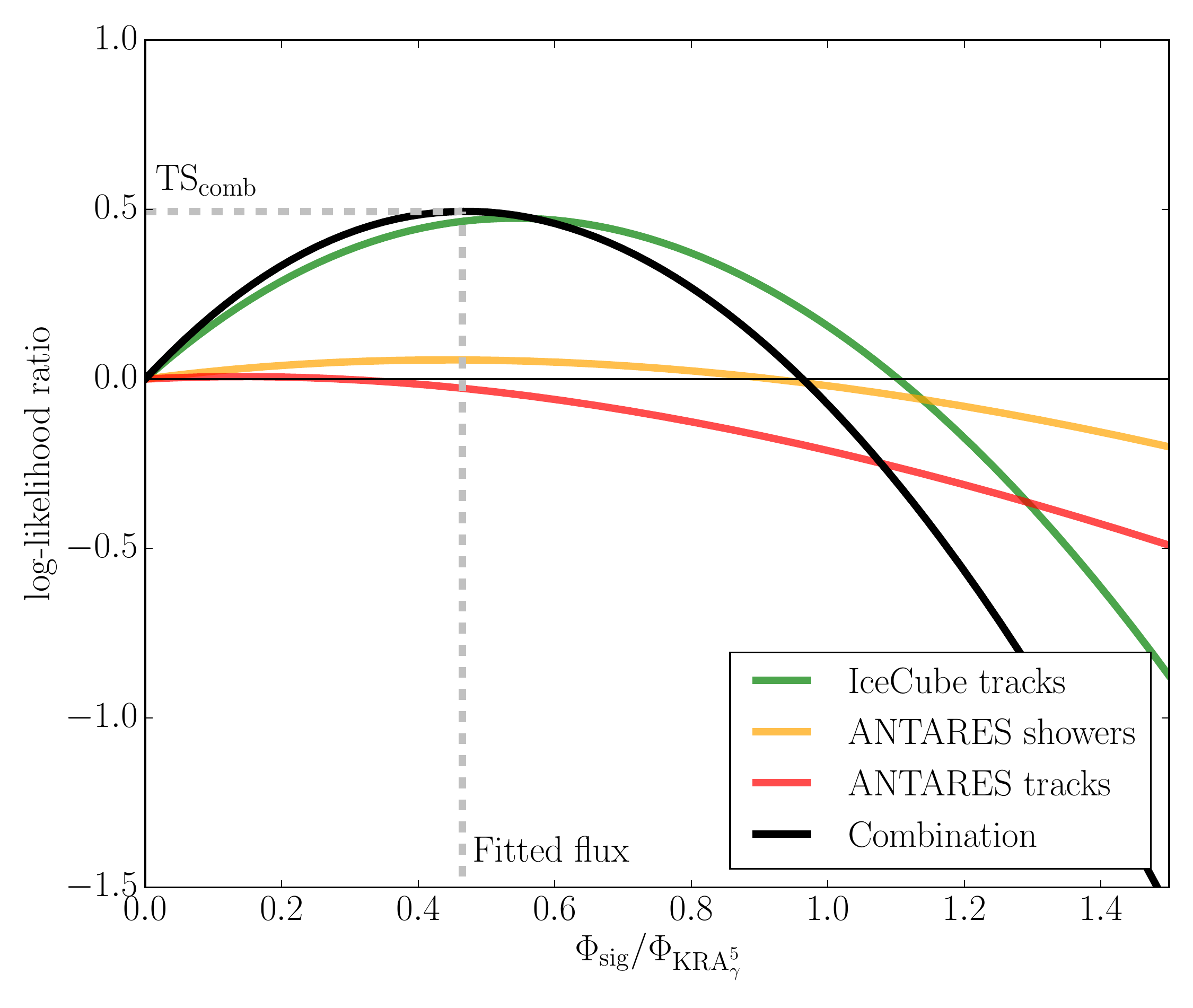}
 \caption{Combination of the log-likelihood ratio curves and fitting of the flux on the combined test statistic. These curves correspond to the unblinded data using the likelihood for the KRA$_\gamma^5$ model.}\label{fig:search_method}
\end{figure}

Then the combined likelihood is simply the product over the per-sample likelihoods.  The combined test statistic is the log-likelihood ratio evaluated for that $\Phi_\textrm{sig}$ which maximizes the combined likelihood:
\begin{equation}\label{eq:ts_comb}
  \textrm{TS}_\textrm{comb} = \max\limits_{\Phi_\textrm{sig}} \left\{\sum\limits_\textrm{sample} \ln\left[\frac{\mathcal{L}_\textrm{sig+bkg} (\Phi_\textrm{sig})} {\mathcal{L}_\textrm{bkg}}\right]\right\},
\end{equation}
where TS$_\textrm{comb}$ is the combined test statistic, $\mathcal{L}_\textrm{bkg} = \mathcal{L}_\textrm{sig+bkg} (\Phi_\textrm{sig}=0)$ is the likelihood to have only background and the sum runs over the event samples.  This is illustrated in Figure~\ref{fig:search_method} with the combined log-likelihood ratio and $\textrm{TS}_\textrm{comb}$ fit for the KRA$_\gamma^5$ model.

\begin{figure*}
 \gridline{
  \fig{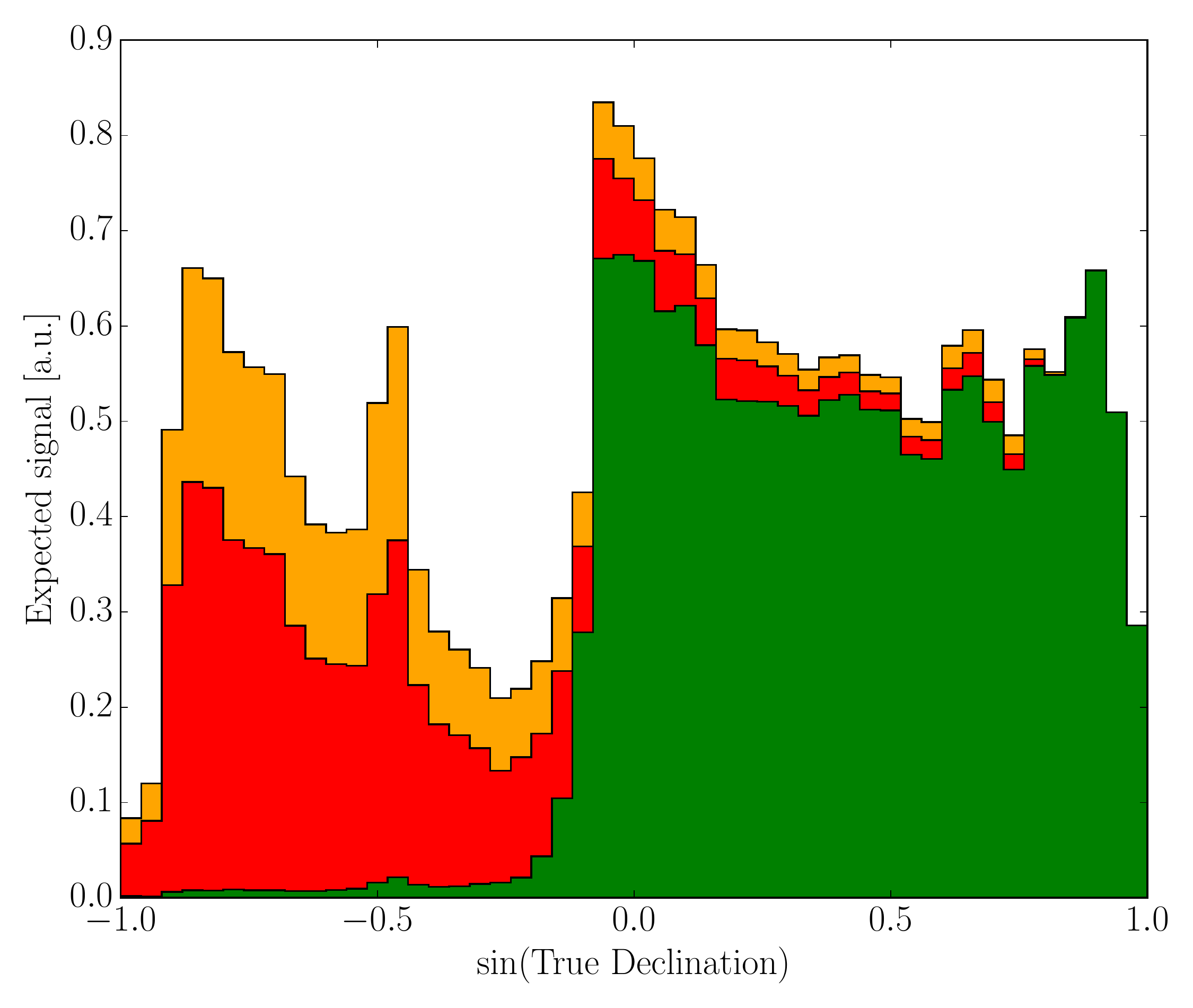}{0.45\textwidth}{(a)}
  \fig{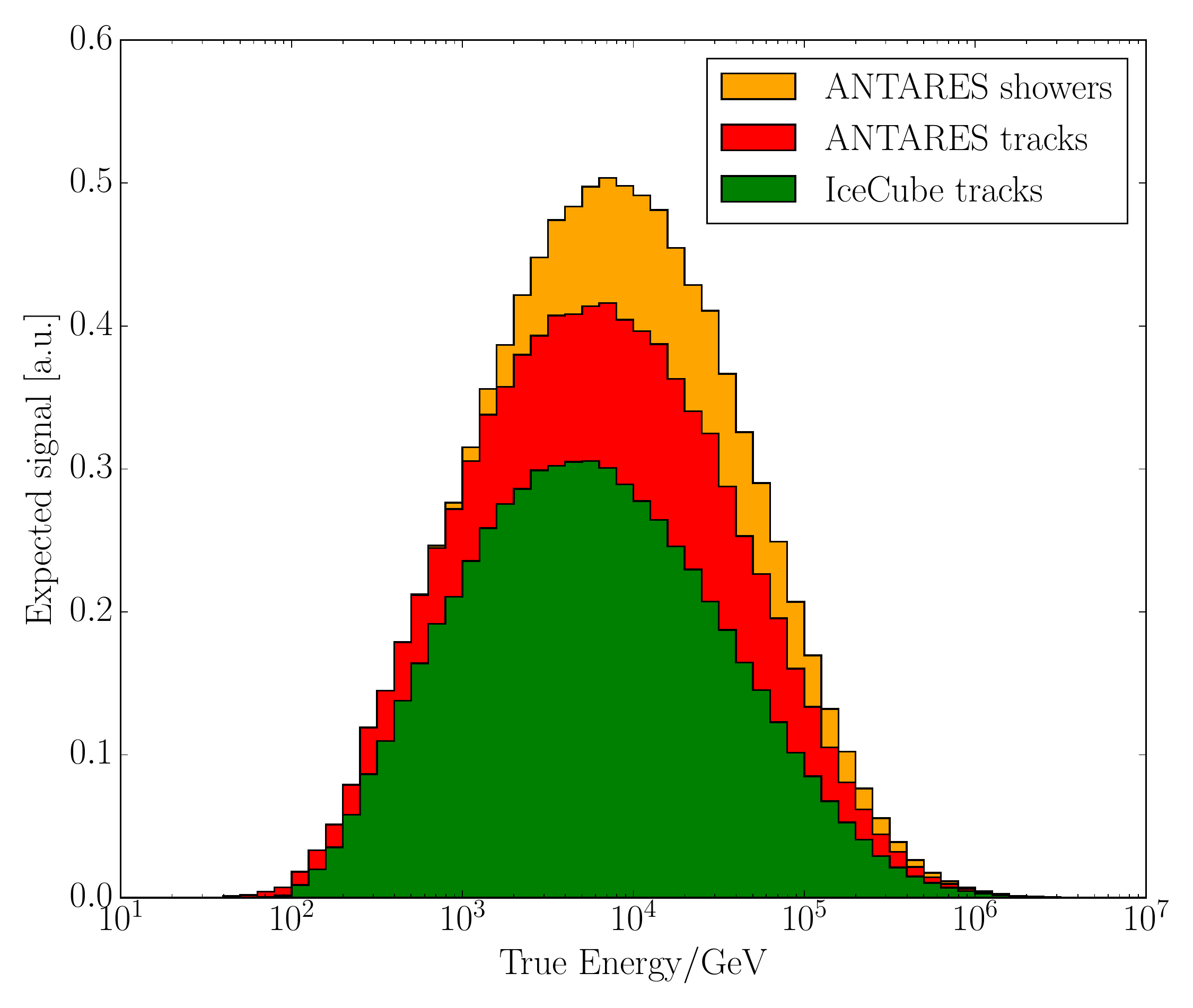}{0.45\textwidth}{(b)}
 }
 \caption{\label{fig:complementarity}Stacked histograms (i.e., every bin shows the fractional contribution of every sample summed on top of each other) of the signal expected from the KRA$_\gamma^5$ model as function of the declination (a) and energy (b) Monte Carlo truth. The colored area of each histogram represents the relative contribution to the sensitivity of this event sample. The relative contribution to the sensitivity is defined as the difference in the sensitivity flux resulting from the addition of a certain event sample divided by the combined sensitivity flux.}
\end{figure*}

The combined and independent sensitivities are summarized in Table~\ref{table:results}. They are defined as the median upper limit\footnote{It is defined as the average upper limit in the ANTARES only analysis~\citep{GalPlane_ANT}. This and the addition of 2016 data account for the difference in the ANTARES sensitivities.}.
The combination is not only a way to exploit more data with different systematics but also an opportunity to benefit from the complementarity of the two detectors.  While IceCube has much higher statistics than ANTARES, we show in Figure~\ref{fig:complementarity}~(a) that ANTARES offers enhanced sensitivity in the southern sky where a larger flux is expected.  This favorable view is coupled with relatively better angular resolution for ANTARES than IceCube.  In Figure~\ref{fig:complementarity}~(b) we show that while IceCube can in principle detect higher energy events compared to ANTARES, the direction-dependent model spectra studied here result in similar energy ranges being tested by both detectors.  Overall, the relative contribution of IceCube to the sensitivity is 61\%; for ANTARES tracks and showers the relative contributions are 25\% and 14\%, respectively.

\section{Results and discussion}
This analysis combines seven years of IceCube tracks and ten years of ANTARES tracks and showers using a likelihood ratio test.  The results are summarized in Table~\ref{table:results}.  Systematic uncertainties on the ANTARES detection efficiency (due to the uncertainty on the acceptance of the ANTARES PMTs) are included in the analysis as in the paper by~\citet{GalPlane_ANT}.  As described by~\citet{IC_PS_7yrs}, systematic uncertainties in the modeling of the Antarctic ice and the optical module efficiency lead to an uncertainty on the IceCube detection efficiency of at most 11\% which is not included here.
\begin{table*}
 \centering
  \caption{\label{table:results}Sensitivities and results of the analysis on the KRA$_\gamma$ models with the 5 and \unit[50]{PeV} cutoffs.}
 \begin{tabular}{ c  c  c  c  c  c  c}\toprule
  \multirow{2}{*}{Energy cutoff} & \multicolumn{3}{c}{Sensitivity $[\Phi_{\textrm{KRA}_\gamma}]$} & Fitted flux & $\it{p}$-value & UL at 90\% CL \\\cmidrule(rl){2-4}
         & Combined & ANTARES & IceCube & $[\Phi_{\textrm{KRA}_\gamma}]$ & [\%] & $[\Phi_{\textrm{KRA}_\gamma}]$ \\\midrule
  \unit[5]{PeV}  & $0.81$   & $1.21$  & $1.14$  & $0.47$  & $29$ & $1.19$ \\
  \unit[50]{PeV} & $0.57$   & $0.94$  & $0.82$  & $0.37$  & $26$ & $0.90$ \\\bottomrule
 \end{tabular}
\end{table*}

The maximum-likelihood estimate yields a non-zero diffuse Galactic neutrino flux for both models with a p-value of 29\% for KRA$_\gamma^{5}$ and 26\% for KRA$_\gamma^{50}$.  Since neither of these results is statistically significant, we place upper-limits on both model normalizations.
The KRA$_\gamma^{50}$ model is constrained at the 90\% confidence level (with an upper limit of $0.9\times \Phi_{\textrm{KRA}_\gamma^{50}}$), while the KRA$_\gamma^5$ model is not yet constrained by our analysis. This was expected as the \unit[50]{PeV} cutoff represents an extreme tuning of the acceleration parameters for the Galactic CRs, while the \unit[5]{PeV} cutoff in light CR can be considered a more reliable case for the Galactic accelerators.

\begin{figure}
 \centering
 \includegraphics[width=0.45\textwidth]{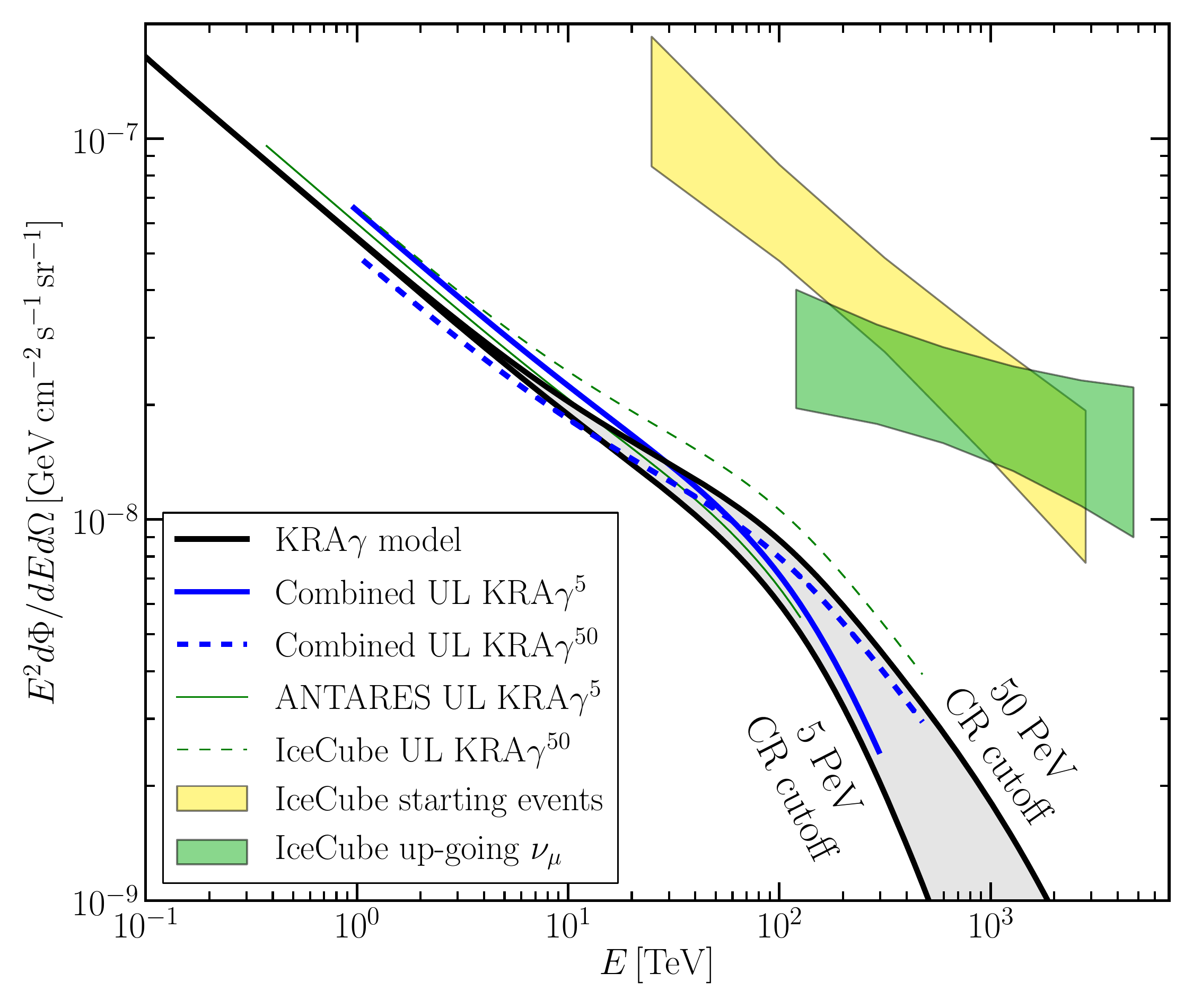}
 \caption{Combined upper limits (UL) at 90\% confidence level (blue lines) on the three-flavor neutrino flux of the KRA$_\gamma$ model with the 5 and \unit[50]{PeV} cutoffs (black lines). The boxes represent the diffuse astrophysical neutrino fluxes measured by IceCube using an isotropic flux template with starting events (yellow) and upgoing tracks (green).}\label{fig:UL}
\end{figure}
Figure~\ref{fig:UL} represents the combined upper limits in comparison to the all-flavor full sky energy spectrum of the KRA$_\gamma$ models as well as the previous IceCube and ANTARES upper limits.  The present upper limit on the \unit[5]{PeV} model is higher than the previously published upper limit for ANTARES alone although the sensitivity is much better. This is due to the overfluctuation observed in the IceCube data sample as well as the difference in the definition of the test statistic.  In the ANTARES standalone analysis it was the sum of the shower and track test statistics, computed independently, instead of computing one test statistic from the combined log-likelihood ratio curve (equation~\ref{eq:ts_comb}).

The results presented here provide for the first time a combined constraint on diffuse Galactic neutrino emission by IceCube and ANTARES.  The limit on the KRA$_\gamma$ model with \unit[50]{PeV} cutoff extends the energy range of the constraint on the model from \unit[10]{GeV} with \Fermi-LAT up to hundreds of TeV.  
Based on the limit on the KRA$_\gamma^{5}$-model, this analysis limits the total flux contribution of diffuse Galactic neutrino emission to the total astrophysical signal reported by \citet{IC_CombinedFlux} to 8.5\%.
In the future, the sensitivity of this analysis can be further improved by including IceCube showers \citep{MESC}. This will allow for a powerful test of the KRA$_\gamma^5$ model, thereby constraining the diffusion mechanisms, the maximal energy injected by supernova remnants and the Galactic gas distributions considered in the model.

\acknowledgements
ANTARES acknowledges the financial support of the funding agencies:
Centre National de la Recherche Scientifique (CNRS), Commissariat \`a
l'\'ener\-gie atomique et aux \'energies alternatives (CEA),
Commission Europ\'eenne (FEDER fund and Marie Curie Program),
Institut Universitaire de France (IUF), IdEx program and UnivEarthS
Labex program at Sorbonne Paris Cit\'e (ANR-10-LABX-0023 and
ANR-11-IDEX-0005-02), Labex OCEVU (ANR-11-LABX-0060) and the
A*MIDEX project (ANR-11-IDEX-0001-02),
R\'egion \^Ile-de-France (DIM-ACAV), R\'egion
Alsace (contrat CPER), R\'egion Provence-Alpes-C\^ote d'Azur,
D\'e\-par\-tement du Var and Ville de La
Seyne-sur-Mer, France;
Bundesministerium f\"ur Bildung und Forschung
(BMBF), Germany;
Istituto Nazionale di Fisica Nucleare (INFN), Italy;
Nederlandse organisatie voor Wetenschappelijk Onderzoek (NWO), the Netherlands;
Council of the President of the Russian Federation for young
scientists and leading scientific schools supporting grants, Russia;
National Authority for Scientific Research (ANCS), Romania;
Mi\-nis\-te\-rio de Econom\'{\i}a y Competitividad (MINECO):
Plan Estatal de Investigaci\'{o}n (refs. FPA2015-65150-C3-1-P, -2-P and -3-P, (MINECO/FEDER)), Severo Ochoa Centre of Excellence and MultiDark Consolider (MINECO), and Prometeo and Grisol\'{i}a programs (Generalitat
Valenciana), Spain;
Ministry of Higher Education, Scientific Research and Professional Training, Morocco.
We also acknowledge the technical support of Ifremer, AIM and Foselev Marine
for the sea operation and the CC-IN2P3 for the computing facilities.

The IceCube Neutrino Observatory acknowledges support from the following agencies:
USA -- U.S. National Science Foundation-Office of Polar Programs,
U.S. National Science Foundation-Physics Division,
Wisconsin Alumni Research Foundation,
Center for High Throughput Computing (CHTC) at the University of Wisconsin-Madison,
Open Science Grid (OSG),
Extreme Science and Engineering Discovery Environment (XSEDE),
U.S. Department of Energy-National Energy Research Scientific Computing Center,
Particle astrophysics research computing center at the University of Maryland,
Institute for Cyber-Enabled Research at Michigan State University,
and Astroparticle physics computational facility at Marquette University;
Belgium -- Funds for Scientific Research (FRS-FNRS and FWO),
FWO Odysseus and Big Science programmes,
and Belgian Federal Science Policy Office (Belspo);
Germany -- Bundesministerium f\"ur Bildung und Forschung (BMBF),
Deutsche Forschungsgemeinschaft (DFG),
Helmholtz Alliance for Astroparticle Physics (HAP),
Initiative and Networking Fund of the Helmholtz Association,
Deutsches Elektronen Synchrotron (DESY),
and High Performance Computing cluster of the RWTH Aachen;
Sweden -- Swedish Research Council,
Swedish Polar Research Secretariat,
Swedish National Infrastructure for Computing (SNIC),
and Knut and Alice Wallenberg Foundation;
Australia -- Australian Research Council;
Canada -- Natural Sciences and Engineering Research Council of Canada,
Calcul Qu\'ebec, Compute Ontario, Canada Foundation for Innovation, WestGrid, and Compute Canada;
Denmark -- Villum Fonden, Danish National Research Foundation (DNRF);
New Zealand -- Marsden Fund;
Japan -- Japan Society for Promotion of Science (JSPS)
and Institute for Global Prominent Research (IGPR) of Chiba University;
Korea -- National Research Foundation of Korea (NRF);
Switzerland -- Swiss National Science Foundation (SNSF).

The IceCube Collaboration designed, constructed and now operates the IceCube Neutrino Observatory. Data processing and calibration, Monte Carlo simulations of the detector and of theoretical models, and data analyses were performed by a large number of collaboration members, who also discussed and approved the scientific results presented here. It was reviewed by the entire collaboration before publication, and all authors approved the final version of the manuscript.

The main authors of this manuscript were Jon Dumm, Timoth{\'e}e Gr{\'e}goire, Christian Haack and Michael Richman.

\end{document}